# On the dimensional indeterminacy of one-wave factor analysis under causal effects

Tyler J. VanderWeele
*Harvard University, Cambridge, MA, U.S.A.*

Charles J. K. Batty
*University of Oxford, Oxford, U.K.*

**Summary.** It is shown, with two sets of indicators that separately load on two distinct factors, independent of one another conditional on the past, that if it is the case that at least one of the factors causally affects the other, then, in many settings, the process will converge to a factor model in which a single factor will suffice to capture the covariance structure among the indicators. Factor analysis with one wave of data can then not distinguish between factor models with a single factor versus those with two factors that are causally related. Therefore, unless causal relations between factors can be ruled out a priori, alleged empirical evidence from one-wave factor analysis for a single factor still leaves open the possibilities of a single factor or of two factors that causally affect one another. The implications for interpreting the factor structure of psychological scales, such as self-report scales for anxiety and depression, or for happiness and purpose, are discussed. The results are further illustrated through simulations to gain insight into the practical implications of the results in more realistic settings prior to the convergence of the processes. Some further generalizations to an arbitrary number of underlying factors are noted.

*Keywords*: Causal effect; covariance; factor analysis; measurement; scale development

## 1. Introduction

Exploratory factor analysis (Thompson, 2004; Comrey and Lee, 2013; Kline, 2014) is frequently used to assess the dimensionality of a set of indicators in a survey. In many cases, the motivation for such an analysis is to allegedly demonstrate that a set of items constitutes a unidimensional scale. Such factor analysis is typically carried out with a single wave of data collected, i.e. a single time point at which all of the items are assessed. Establishing the unidimensionality of a scale is often viewed as an important part of scale development (DeVellis, 2016), to be carried about before the scale is employed in longitudinal data collection efforts.

It is well-known that one cannot typically assess causal relations with a single wave of data in which data on all variables is collected at the same time (Morgan and Winship, 2015; VanderWeele, 2015; Hernán and Robins, 2020). However, the implications of this fact for the psychometric evaluation of scales has largely been ignored. If, for example, there are two underlying factors that explain a set of survey item responses or indicators, and if these factors are causally related, it will not be possible, with a single wave of data, to assess causal relations between them. Unfortunately, this has rather serious consequences for attempts to assess factor dimensionality with a single wave of data. Specifically, with a single wave of data, it will not be possible to distinguish associations arising from causal relations between the factors from those allegedly arising from conceptual relations among the indicators. The present paper formalizes this intuition and discusses the implications for the practice of factor analysis.

## 2. Factor Analysis with Two Causally Related Factors

Consider a standard factor analytic model (Thompson, 2004; Comrey and Lee, 2013; Kline, 2014), with two sets of survey item responses, $(Y_1^t, \ldots, Y_w^t)$ and $(Y_{w+1}^t, \ldots, Y_p^t)$, measured at time $t$, that separately load on two distinct factors, $\eta_1^t$ and $\eta_2^t$, independent of one another conditional on past values of the latent factors, $\eta_1^{t-1}$ and $\eta_2^{t-1}$. Suppose, however, that over time at least one of the factors causally affects the other, which then gives rise to what is sometimes called a dynamic factor model or dynamic structural equation model (Stock and Watson, 2011; Shumway and Stoffer, 2017; Asparouhov *et al*., 2018), as illustrated in Figure 1.

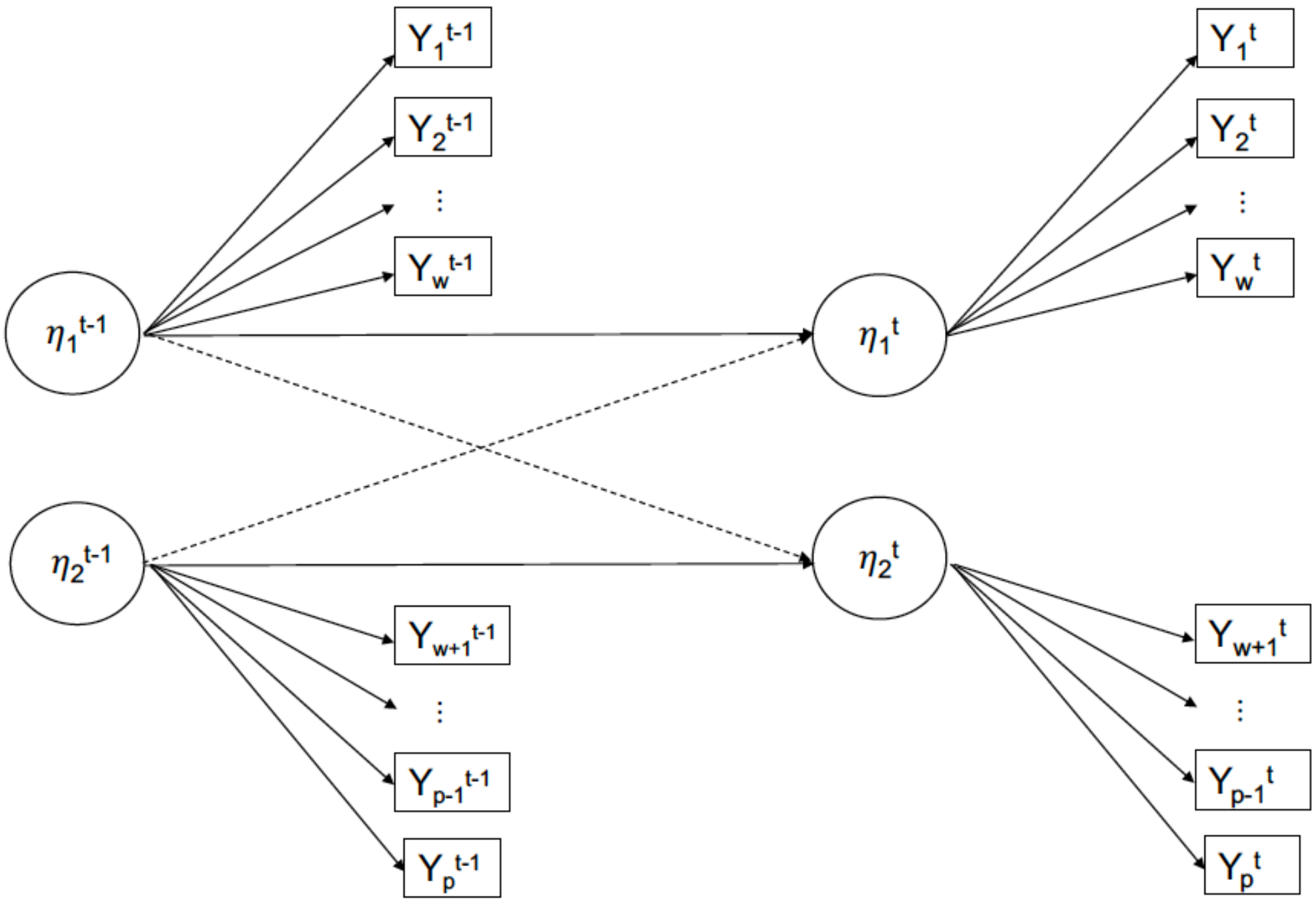

Figure 1. Causal effects of two latent factors on each other over time

More generally, if we let $Y^t = (Y_1^t, \ldots, Y_p^t)'$ denote a set of item responses or indicators measured at time $t$, and $\eta^t = (\eta_1^t, \ldots, \eta_m^t)'$ be a set of $m$ latent factors at time $t$, then the standard factor analytic model with independent errors is given by:

$$Y^t = \Lambda\eta^t + \varepsilon$$

where $\Lambda$ is an $p \times m$ matrix and $\varepsilon$ is an $p \times 1$ vector of independent normally distributed random variables. For simplicity, assume, as is often the case, that the variables $Y^t$ have been standardized to have mean 0 and variance 1. Exploratory factor analysis (Thompson, 2004; Comrey and Lee, 2013; Kline, 2014) attempts to draw conclusions about the dimensionality of $\eta^t$ using the observed data $Y^t$. While this is in some sense a very specific model, it is the model that is effectively employed in thousands of applied papers on scale development (Comrey and Lee, 2013; DeVellis, 2016), almost always with data at a single wave i.e. at a single time point $t$.

Now suppose that the latent factors $\eta^t$ can change over time and that the components may be causally related to each other but that the entire vector of latent variables $\eta^t$ follows a Markov process so that:

$$\eta^t = B\eta^{t-1} + W^t$$

where $B$ is an $m\times m$ matrix with the *i-j* entry representing the causal effect on factor *i* at time *t* of factor *j* at time *t-1*, and where $W^t$ is a $m\times 1$ vector of random errors. We can consider the behaviour of the process constituted by the set of indicators $Y^t = (Y_1^t, \dots, Y_p^t)'$ over time. As time passes, we have the following result concerning the convergence of this process.

**Theorem 1** *Suppose indicators $Y^t = (Y_1^t, \dots, Y_p^t)'$ at time t are related to a set of m latent factors $\eta^t = (\eta_1^t, \dots, \eta_m^t)'$ by:*

$$Y^t = \Lambda\eta^t + \varepsilon$$

*and that $\eta^t = B\eta^{t-1} + W^t$ where the variables $W^t$ are independently normally distributed with potentially distinct parameters at each t. Let $B = QDQ^{-1}$ be Jordan decomposition of B where D is an $m\times m$ matrix of Jordan normal form. Suppose it is the case that (i) as $t \to \infty$, $Y^t$ converges in distribution to some random variable $Y^*$, (ii) B is invertible and (iii) the random variables $W^t$ decay sufficiently quickly such that as $t \to \infty$, $(\sum_{k=1}^{t} B^{-k} W^k)$ converges in distribution to some normally distributed variable $W^*$, then $Y^*$ will follow a factor model:*

$$Y^* = \Lambda^*\eta^* + \varepsilon$$

*with dim($\eta^*$)= rank($D^*$) where $D^* = \lim_{t\to\infty} D^t$.*

**Proof**. Since $\eta^t = B\eta^{t-1} + W^t$ we then have that

$$\begin{aligned} Y^t &= \Lambda\eta^t + \varepsilon = \Lambda(B\eta^{t-1} + W^t) + \varepsilon \\ &= \Lambda B\eta^{t-1} + \Lambda W^t + \varepsilon \\ &= \Lambda B(B\eta^{t-2} + W^{t-1}) + \Lambda W^t + \varepsilon \\ &= \Lambda B^2\eta^{t-2} + \Lambda \mathrm{B} W^{t-1} + \Lambda W^t + \varepsilon \end{aligned}$$

and by iteration:

$$Y^t = \Lambda B^t\eta^0 + \Lambda\left(\sum_{k=0}^{t-1} \mathrm{B}^k W^{t-k}\right) + \varepsilon$$

For the process $Y^t$ to converge in distribution as $t \to \infty$ to some random variable $Y^*$, we must have that the matrix $B^t$ converges as $t \to \infty$. Let $B = QDQ^{-1}$ denote the Jordan decomposition of $B$ for some $m\times m$ invertible matrix $Q$ and where $D$ is an $m\times m$ matrix of Jordan normal form (Meyer, 2000). It then follows that $B^t = QD^tQ^{-1}$ and $B^t$ will converge if and only if $D^t$ converges. By Theorem 1 of Oldenburger (1940), for $D^t$ to converge as $t \to \infty$, it must have, as its limit, subject to permutation of indices, a matrix of the form:

$$D^* = \lim_{t\to\infty} D^t = \begin{pmatrix} I & 0 \\ 0 & 0 \end{pmatrix}$$

where either the identity or the zeros along the diagonal may possibly be absent. Define $B^* = lim_{t\to\infty}B^t$ then

$$B^* = \lim_{t\to\infty} B^t = \lim_{t\to\infty} QD^tQ^{-1} = QD^*Q^{-1}.$$

The random variables $W^t$ decay sufficiently quickly over time such that as $t \to \infty$ $(\sum_{k=1}^{t} \mathrm{B}^{-k}\, W^k)$ converges in distribution to some normally distributed variable $W^*$ and thus provided, as above, $B^t$ converges to some matrix $B^*$ as $t \to \infty$, we would have

$$\left(\sum_{k=0}^{t-1} \mathrm{B}^k\, W^{t-k}\right) = \mathrm{B}^t \left(\sum_{k=1}^{t} \mathrm{B}^{-k}\, W^k\right) \xrightarrow{d} B^*W^*.$$

We then have that $Y^t = \Lambda B^t \eta^0 + \Lambda(\sum_{k=0}^{t-1} \mathrm{B}^k\, W^{t-k}) + \varepsilon$ converges in distribution to a random variable $Y^*$ given by:

$$Y^* = \Lambda B^* \eta^0 + \Lambda B^* W^* + \varepsilon$$
$$= \Lambda Q D^* Q^{-1} \eta^0 + \Lambda Q D^* Q^{-1} W^* + \varepsilon.$$

Define $\Lambda^* = \Lambda Q, \eta^* = D^* Q^{-1}(\eta^0 + W^*)$. In equilibrium as $t \to \infty$, the factor model could thus be written as:

$$Y^* = \Lambda^* \eta^* + \varepsilon$$

and since $Q$ is invertible, the dimensionality of $\eta^*$ will be rank($D^*$). This completes the proof.

Theorem 1 implies that, under the conditions given in the Theorem, the process constituted by the set of indicators $Y^t = (Y_1^t, \dots, Y_p^t)'$ will, over time, converge to a factor model

$$Y^* = \Lambda^* \eta^* + \varepsilon$$

where the dimensionality of the factors $\eta^*$ depends on the matrix *D* in the Jordan normal form of the matrix *B* denoting the causal effects of each of the factors on the others. The result in fact has particularly striking consequences when there are only two factors $\eta^t = (\eta_1^t, \eta_2^t)$, as stated in the following Corollary.

**Corollary** *Under the conditions of Theorem 1 if dim($\eta^t$)=2 then as $t \to \infty$ the resulting factor model:*

$$Y^* = \Lambda^* \eta^* + \varepsilon$$

*is such that dim($\eta^*$)= 2 if and only if B=I.*

**Proof**. With two latent factors, $\eta^t = (\eta_1^t, \eta_2^t)$, $B$, $D$, and $D^*$ will all be 2×2 matrices. Recall that *D* is the Jordan normal form of *B*. If *D* has the form

$$\begin{pmatrix} \lambda & 1 \\ 0 & \lambda \end{pmatrix}$$

then $D^t$ will only converge if $|\lambda| < 1$ (Oldenburger, 1940). However, in that case, $D^* = lim_{t\to\infty} D^t = 0$. If *D* has the form

$$\begin{pmatrix} \lambda_1 & 0 \\ 0 & \lambda_2 \end{pmatrix}$$

then $D^t$ will only converge if $-1 < \lambda_1 \leq 1$ and $-1 < \lambda_2 \leq 1$ (Oldenburger, 1940). If one of $\lambda_1$ or $\lambda_2$ is less than in absolute value 1 and the other is equal to 1 then rank($D^*$)= $rank(lim_{t\to\infty} D^t) = 1$. If both are less than 1 in absolute value then rank($D^*$)=0. Thus the only way we can have rank($D^*$)=2 is if $\lambda_1$=$\lambda_2$=1 in which case, $D$=$I$ and $B = QDQ^{-1} = QQ^{-1} = I$. This completes the proof.

The corollary states that for two latent factors $\eta^t = (\eta_1^t, \eta_2^t)$, under the condition of Theorem 1, the only way for the process constituted by the set of indicators $Y^t = (Y_1^t, \dots, Y_p^t)'$ to converge over time to a factor model which itself has two factors, so that dim($\eta^*$)= 2, is if the matrix $B$ denoting the causal effects of each of the factors on the others is the identity matrix i.e. if neither $\eta_1^{t-1}$ affects $\eta_2^t$, nor $\eta_2^{t-1}$ affects $\eta_1^t$ so that there are no causal effects of the factors on one another.

We are thus left with the conclusion that, in equilibrium as $t \to \infty$, either there are no causal effects of the factors on one another, or if there were, then a factor model with a single factor will be sufficient. This latter result is in some sense analogous to the Perron-Frobenius theorem for time-homogeneous Markov chains with finite state spaces (cf. Serfozo, 2009) for which if the Markov chain is irreducible and aperiodic, then the transition matrix raised to the k[th] power will converge to a rank-one matrix. In the present dynamic factor model under consideration, $\eta^t$ is itself a Markov chain on a measurable state space, although $Y^t$ is not. Note that in the proof of Theorem 1, although the factors constituted by the components of the vector $\eta^* = D^* Q^{-1}(\eta^0 + W^*)$ may be correlated, if $rank(D^*) = 1$ then we will still be left with only a single factor.

Stated another way, Corollary 1 indicates that causal effects of one latent factor on another imply that, in equilibrium, if only a single wave of data is collected, a factor model with one factor will suffice, even if the true underlying structures are such that there are two causally related factors. The implications of this result for the current practice of factor analysis are unsettling. Efforts are often made during scale development to demonstrate unidimensionality of a set of item responses using factor analysis with one wave of data (DeVellis, 2016). If factor analysis provides evidence that a single factor is sufficient to explain most of the covariance in the item responses, this is generally deemed satisfactory. However, the argument above indicates that this is also exactly the empirical result that one might expect, with one-wave factor analysis, if there were in fact two distinct latent factors that causally affected each other over time. With a one-wave factor analysis, if there is evidence for more than one factor, then this is genuine evidence *against* unidimensionality. But if a one-wave factor analysis suggests only one factor, then we cannot distinguish between the possibilities of two factors with causal effects over time versus a single factor. The only way we could establish unidimensionality in this case would be if we could rule out, a priori, that, if there were two or more factors, then they definitively did not causally affect one another. But it is difficult to imagine circumstances in which we were uncertain, on conceptual grounds, about the number of factors, but knew that, if more than one existed, then they were causally unrelated. We are left with the conclusion that, in many circumstances, alleged empirical evidence for a single factor from one-wave factor analysis in fact essentially cannot rule out the possibility of the presence of two factors with causal effects on one another.

A number of potential objections to the conclusions of the analysis above concerning current factor analytic practices might be put forward. First, Figure 1 considers only a relatively simple factor model with each indicator loading only on one of the two factors, and with independent errors. In some sense though, this is an ideal case, when one might most expect to be able to discern two separate factors from a one-wave factor analysis. And even in this ideal case, one could not in equilibrium distinguish the models with a single factor versus two factors that were causally related. However, Theorem 1 and its Corollary did not in fact rely on indicators loading on only a single factor or on independent errors. Even under these more complex structures, if at least one factor causally affects the other then the argument above shows that, in equilibrium, a factor model with one factor will suffice.

Second, the argument above also imposed certain assumptions on the matrix $B$ and random errors $W^t$. Other, or more general, specifications could be considered. However, this case does suffice to demonstrate how causal effects of the factors on one another, can, in equilibrium, lead to a reduction in the number of factors that suffice for a factor model to account for associations among item responses. The argument did require a decaying structure of the errors so as to obtain convergence. Such a model of the errors over time may not always be realistic. However, in settings in which exogenous sources of variation are more common earlier in life and the variables or states are more stable later in life, such assumptions on the decaying nature of the error terms may be a reasonable approximation. Such may be the case with, for example, education and say the quantile of total wealth. Each of these may causally affect the other, and while either of these can in principle change as time goes on, the likelihood of large increases in education, or in quantile of total wealth, diminishes substantially by mid-life. Likewise, if lifelong trajectories of anxiety and depression are more powerfully shaped by life circumstances, experiences, and therapies in childhood, adolescence and early adulthood, and typically over time become somewhat more stable by mid-life at either relatively low high or relatively high levels psychological distress, still subject to variation but less so, then once again a decaying error structure may be a reasonable approximation.

Third, one might dispute that equilibrium is ever achieved and object that the notion of convergence as $t \to \infty$, is a theoretical abstraction. However, the limit argument above does imply that, if at least one of two factors causally affects the other so that $B$ is not the identity matrix, then one can always find a finite number of time steps $k$, such that $D^k$ is within any given arbitrarily small deviation from being either 0 or the matrix

$$\begin{pmatrix} 1 & 0 \\ 0 & 0 \end{pmatrix}$$

and thus for there being only very slight deviation from a single factor explaining the set of items at time $k$. Moreover, in practice, when investigators use factor analysis to determine the number of factors they will often ignore factor loadings below 0.3 and indeed the guidance given in textbooks on exploratory factor analysis is to do precisely this (Field, 2013). If relatively low factor loadings are indeed ignored, then in principle only a very small number of steps may be sufficient for an investigator to erroneously conclude in practice that a single factor were present. We illustrate this below in the simulation studies. Thus, once again, if there are causal effects of one factor on another and the process $\eta^t$ has proceeded through even a small number of steps, then a one-wave factor analysis can in practice, long before convergence is attained or even approximated, erroneously suggest a single factor based on current factor analysis practices.

It is possible that sufficiently prior to convergence being obtained, factor analyses with one wave of data could be employed to uncover aspects of the underlying processes. For example, under the data-generating structure in Figure 1, prior to convergence, exploratory factor analysis, with a sufficiently large sample size, allowing for correlated/oblique factors (e.g. using varimax rotation, cf. Thompson, 2004; Comrey and Lee, 2013; Kline, 2014) could uncover the fact that the two sets of items loaded on two separate factors. However, an analyst who instead employed independent/orthogonal factors would attain an equally good fit with two independent factors and all items loading on both factors. The factor model, even sufficiently prior to convergence, is unidentified from one wave of data and so it is impossible to distinguish between the two solutions. In some sense, each analyst would correctly identify some aspect of the underlying processes: the first analyst would correctly identify distinct sets of factor loadings, and the second analyst would correctly identify the potential independence of the factors (though, at a given wave, they are only independent of each other conditional on the past). However, it is not possible to fully identify the correct structure with only one wave of data. Again, the causal relationships between the factors renders such identification impossible with one wave of data.

More generally, regardless of the equilibrium argument, and the specification of the error terms, it should be clear even from Figure 1 itself, without any further algebraic derivation, that if there are two factors that causally affect one another, and a factor model is fit with a single wave of data, the covariance amongst the indicators will arise both from the underlying factor for each indicator, and from the causal effect of one factor on the other. The model with factor loadings for each indicator and with causal effects across factors is effectively unidentified with a single wave of data. There is no way to distinguish the two sources of covariance with one wave of data. If there were no causal effects of $\eta_1^{t-1}$ on $\eta_2^t$, nor of $\eta_2^{t-1}$ on $\eta_1^t$, then items $Y_1^t$ and $Y_{r+1}^t$ would be statistically independent. In the presence of a causal effect of $\eta_2^{t-1}$ on $\eta_1^t$, they will be statistically dependent. It will not be possible to distinguish between causal relations among the factors and the allegedly conceptual relations arising from underlying latent factors using just one wave of data.

## 3. Example from Prior Literature: Anxiety and Depression

Feldman (1993) examined self-report scales for anxiety and depression and employed factor analysis to assess whether these self-report anxiety and depression scales measure distinct constructs. Based on results from factor analysis with one wave of data she concludes, "These analyses provide evidence that anxiety and depression self-report scales do not measure discriminant mood constructs and may therefore be better thought of as measures of general negative mood rather than as measures of anxiety and depression per se." Similar conclusions were drawn by Norton *et al.* (2013) in a meta-confirmatory factor analysis using data from 28 samples concerning the Hospital Anxiety and Depression Scale. They conclude that "Due to the presence of a strong general factor, [the Hospital Anxiety and Depression Scale] does not provide good separation between symptoms of anxiety and depression. We recommend it is best used as a measure of general distress."

However, these conclusions ignore the role that causal relationships between the phenomena of depression and anxiety may play in these factor analytic approaches. There is in fact evidence from numerous longitudinal studies that the experience of anxiety renders subsequent depression more likely, and likewise that the experience of depression renders subsequent anxiety more likely (Jacobson and Newman, 2017). It is likely that each has causal effects on the other. This of course has implications for the interpretation of factor analyses such as those of Feldman (1993) and Norton *et al.* (2013). If the experience of anxiety causes depression, and the experience of depression causes anxiety, or even if just one of these two causal relations held, then even if it were the case that the anxiety and depression items loaded on distinct anxiety and depression factors, from the results above, one might still anticipate, from a one-wave factor analysis, evidence for a factor structure with only a single factor. The results of a one-wave factor analysis, such as those of Feldman (1993) and Norton *et al.* (2013), are exactly what one might expect, even if there were two distinct causally related factors. Allowing for the possibility of causally related factors, and in this case there is good reason to expect that possibility, the results of Feldman (1993) and Norton *et al.* (2013) then cannot adequately distinguish between the possibility of two separate factors with causal effects versus a single factor. The basic emotions of sadness and fear, underlying depressive and anxiety disorders respectively, are arguably clearly conceptually distinct. It may well be the case that the only reason the analyses of Feldman (1993) and Norton *et al.* (2013) supposedly indicate that the two cannot be separated is because there are causal relations which their analyses cannot assess because they use only one wave of data.

## 4. Data Analysis Example: Happiness and Purpose

Weziak-Bialowolska et al. (2017) considered data from a Health and Well-Being Survey including various well-being indicators for happiness, health, purpose, character, relationships and financial security. Using factor analysis with one wave of data, they presented evidence for distinct well-being factors for health, character, relationships, and financial security. However, the factor analyses did not indicate separate factors for happiness and purpose in life, but suggested these two as a single factor. It is possible is that the failure to distinguish these factors arises in part from causal effects of purpose in life on happiness, and possibly also from effects of happiness on purpose in life. There is some evidence, from longitudinal studies, for effects in both of these directions, and the evidence for effects of purpose on happiness and life satisfaction is perhaps especially pronounced (Kim et al., 2021, 2022).

We revisit here this issue of distinguishing between happiness and purpose using data from a Wellbeing Assessment in year 2019 of 1,209 employees at a large insurance company (Weziak-Bialowolska et al., 2021). Specifically, we will consider data on three items for happiness, namely: "How satisfied are you with life as a whole these days?", "How happy have you felt during the last 7 days?" and "I expect more good things in my life than bad," each scored 0-10; and also three items for purpose: "To what extent do you feel the things you do in your life are worthwhile?", "I understand my purpose in life", and "I am pursuing what is most important to me in my life," again each scored 0-10. In fitting a factor model using varimax rotation (Thompson, 2004; Comrey and Lee, 2013; Kline, 2014) with two factors, the factor loadings on the first factor are $(0.812, 0.836, 0.539, 0.625, 0.355, 0.403)$ and the factor loadings on the second factor are $(0.397, 0.335, 0.515, 0.536, 0.851, 0.770)$. The first purpose item has a larger factor loading on what is supposedly the happiness factor than on the purpose factor, and this factor loading on the happiness factor for this purpose item is also larger than the happiness factor loading for the third happiness item. The third happiness item moreover has a substantial factor loading on the purpose factor, of roughly the same magnitude as its loading on the happiness factor. Furthermore all six of the items have factor loadings above 0.3 on both factors. This one-wave factor analyses may, however, in part be confounded by the effects that each of past happiness and past purpose-in-life have on the present values of other.

Fortunately, in the case of this well-being survey, data is also available on the same items, one year prior, in 2018. To attempt to control for potential causal effects of purpose and happiness, stratification on past values of purpose and happiness can be carried out. Given prior evidence for the especially strong effects of purpose on happiness, attention could be restricted to the subgroup for which levels of purpose in 2018 were below the median and levels of happiness in 2018 were already above the median. Fitting again a factor model using varimax rotation with two factors with the 2019 data, with restriction to this stratum to attempt to control for confounding for causal effects of past purpose on happiness, the factor loadings on the first factor then becomes $(0.867, 0.804, 0.459, 0.449, 0.194, 0.172)$ and the factor loadings on the second factor are $(0, 0.211, 0.249, 0.399, 0.751, 0.865)$. While the separation of the factors is not perfect, it is considerably better. The first purpose item no longer has a stronger factor loading on the happiness factor than the third happiness item; all of the happiness items have factor loadings on the purpose factor that are below 0.3; and all of the purpose items except the first now also have factor loadings on the happiness factor that are below 0.3. Although such crude stratification by past values of happiness and purpose only partially controls for the causal effects of purpose on happiness, we see that even with this relatively crude form of control, there is greater separation between the happiness and purpose factors.

## 5. Simulations

In this section we illustrate the results further through simulations to gain insight into the implications of the results in more realistic settings in which only a finite and possibly small number of time periods or steps have taken place so that convergence is not yet achieved. Consider again, the dynamic factor model with $Y^t = (Y_1^t, \ldots, Y_p^t)'$ denoting a set of indicators measured at time $t$, and $\eta^t = (\eta_1^t, \ldots, \eta_m^t)'$ a set of $m$ latent factors at time $t$, with:

$$Y^t = \Lambda\eta^t + \varepsilon$$

where $\Lambda$ is an $p\times m$ matrix and $\varepsilon$ is an $p\times 1$ vector of independent normally distributed random variables. For simplicity, we will consider the setting with $m=2$ and with 3 indicators per latent variable so that $p=3$ and we will assume the variables $Y^t$ have been standardized to have mean 0 and variance 1. Suppose again, as above, that:

$$\eta^t = B\eta^{t-1} + W^t$$

where $B$ is an $2\times 2$ matrix with the *i-j* entry representing the causal effect on factor $i$ at time $t$ of factor $j$ at time $t$-*1*, and where $W^t$ is a $2\times 1$ vector of random errors.

Consider a setting in which at each time $t$ the first three items load only on the first factor and the second three items load only on the second factor, with relatively strong factor loadings given by $\Lambda_{\cdot 1} = (0.7, 0.7, 0.8, 0, 0, 0)'$ and $\Lambda_{\cdot 2} = (0, 0, 0, 0.7, 0.8, 0.6)'$.This will be a rather ideal scenario as the "cross-loadings" are all 0 and the other factor loading are relatively large in magnitude. We will see, however, that, even in this scenario, the processes begin to converge relatively quickly to a model that can be approximated by a single factor. Consider the setting with

$$B = \begin{pmatrix} 0.8 & \gamma \\ \gamma & 0.8 \end{pmatrix}$$

so that $\gamma$ denotes the magnitude of the causal effect of each factor on the other. Let each of the two components of $W^t$ be normally distributed with mean 0 and standard deviation $\sigma\delta^t$ so that $\delta$ denotes how quickly the random variation in the process governing the latent factors $\eta^t$ decays over time.

We will first illustrate the results above with a particular set of parameter values and then consider, over a broad range of the parameter values of $\gamma$ and $\delta$ and sample size $N$, how quickly the process becomes effectively indistinguishable from a one-factor model when using only one wave of data.

Consider first the setting with $\gamma=0.35$, $\sigma = 0.3$ and $\delta = 0.9$, with sample size $N$=1,000. Note that a standardized effect size for the effects of one factor on another of $\gamma=0.35$ would be somewhat *smaller* than the estimated effects of depression and anxiety on each other (Jacobson and Newman, 2017) and perhaps slightly larger than the effect of happiness on purpose (Kim et al., 2021) and similar to or slightly smaller than the effect of purpose on happiness (Kim et al., 2022). Code for the simulations that follows is available in the Online supplement. With an initial simulated dataset (t=0), the estimates of the factor loadings from a factor model using varimax rotation are $\Lambda_{\cdot 1} = (0.73, 0.69, 0.83, 0, 0, 0)'$ and $\Lambda_{\cdot 2} = (0, 0, 0, 0.70, 0.83, 0.57)'$ and a $\chi^2$test with 9 degrees of freedom under the null hypothesis that one latent factor (with only 6 factor loading parameters $\Lambda_{\cdot 1}$) is sufficient, as compared with an unconstrained model for the correlation matrix with 15 parameters, gives a p-value of $9.1 \times 10^{-145}$ . After only three time steps (t=3), the factor loading estimates, if only a single wave of data at t=3 is used, are $\Lambda_{\cdot 1} = (0.65, 0.64, 0.64, 0.39, 0.53, 0.43)'$ and $\Lambda_{\cdot 2} = (0.38, 0.42, 0.51, 0.67, 0.63, 0.52)'$ and the distinct factors are much less clearly distinguishable from one another. However, it is still the case that a $\chi^2_9$ test statistic under the null hypothesis that one latent factor is sufficient gives a p-value of 0.0039. By t=5, the factor loading estimates are rotation $\Lambda_{\cdot 1}$= $(0.73, 0.71, 0.77, 0.69, 0.54, 0.62)'$ and $\Lambda_{\cdot 2}$ = $(0.41, 0.44, 0.46, 0.46, 0.84, 0.42)'$ and the $\chi^2_9$ test statistic under the null hypothesis that one latent factor is sufficient now gives a p-value of 0.57. It is effectively no longer possible to distinguish the two factors.

To gain further insight into how quickly the two factors become indistinguishable, we consider a range of scenarios in which we vary the magnitude of the effect of the factors on each other, $\gamma$, the rate $\delta$ at which the random variation in the process governing the latent factors decays over time, and the sample size *N*. We will consider $\gamma = (0.1, 0.35, 0.5)$ corresponding to small, moderate, and large effects sizes of the factors on one another; $\delta = (0.65, 0.8, 0.9)$ corresponding to fast, moderate, and relatively slow rates of decay at which the random variation in the process governing the latent factors decreases, and $N = (300, 1000, 3000)$ corresponding to modest, moderate, and relatively large sample sizes as employed in factor analysis. For each parameter setting, 500 simulated datasets are created. For each set of parameters, we report in Table 1 the mean, standard deviation, minimum maximum, median, and 25$^{th}$ and 75$^{th}$ quantiles of number of iterations across the 500 datasets before the $\chi^2_9$ test statistic under the null hypothesis that one latent factor is sufficient rises above the p=0.05 threshold.

As can be seen from Table 1, in a relatively small number of iterations it effectively becomes no longer possible to distinguish the two factors. Each of the parameters that is varied clearly matters. As would be expected, larger cross-factor effects $\gamma$, faster decay of the random variation in the process for the underlying latent variables $\delta$, and smaller sample size *N* all result in the factors effectively becoming indistinguishable in fewer iterations. Even with small cross-factor effects ($\gamma$=0.1), slow decay of the random variation in the process for the underlying latent factor ($\delta = 0.9$) and a large sample size (*N*=3000), only about 10 iterations are needed before the factors effectively become indistinguishable. When the cross-factor effects are moderate or large ($\gamma$=0.35 or $\gamma$=0.5) and the decay of the random variation in the process for the latent factors is moderate or fast ($\delta = 0.8$ or $\delta = 0.9$), then often only 2 or 3 iterations is needed before the factors effectively become indistinguishable even when the sample size is large. As can be seen from Table 1 there is some variation across the 500 datasets with regard to the number of iterations before the factors become effectively indistinguishable. The parameter that most strongly affects the number of iterations before the factors become effectively indistinguishable appears to be the rate of decay of the random variation in the process for the underlying latent factors. When $\delta$=0.9, the standard deviation in the number of required iteration is considerable higher across the various other parameter settings. The mean and median number of required iterations is still often quite modest, but the standard deviation and the maximum of the number of required iterations is notably larger when $\delta = 0.9$. When the cross-factor effects were large ($\gamma$=0.5) with a large sample size (*N*=3000), and a slow decay rate ($\delta = 0.9$), in one of the 500 simulated datasets, the $\chi^2$test had not rejected the null hypothesis of one factor even after 40 iterations.

**TABLE 1** Means, standard deviations, minimum, maximum, median, and 25[th] and 75[th] quantiles of number of iterations across the 500 datasets before the two distinct factors become effectively indistinguishable based on a $\chi_9^2$ test for the null hypothesis of one factor across scenarios defined by cross-factor effects ($\gamma$), the rate of decay of random variation in the process underlying the latent factor ($\delta$), and the sample size (*N*)

| N | $\gamma$ | $\delta$ | Mean | St. Dev. | Min | 25% | Median | 75% | Max |
|---|---|---|---|---|---|---|---|---|---|
| 300 | 0.1 | 0.9 | 5.56 | 1.05 | 3 | 5 | 5 | 6 | 10 |
| 300 | 0.1 | 0.8 | 4.66 | 0.78 | 3 | 4 | 5 | 5 | 7 |
| 300 | 0.1 | 0.65 | 4.19 | 0.72 | 3 | 4 | 4 | 5 | 6 |
| 300 | 0.35 | 0.9 | 2.94 | 0.78 | 2 | 2 | 3 | 3 | 6 |
| 300 | 0.35 | 0.8 | 2.50 | 0.63 | 2 | 2 | 2 | 3 | 5 |
| 300 | 0.35 | 0.65 | 2.26 | 0.49 | 2 | 2 | 2 | 2 | 5 |
| 300 | 0.5 | 0.9 | 2.30 | 0.70 | 1 | 2 | 2 | 3 | 6 |
| 300 | 0.5 | 0.8 | 2.03 | 0.55 | 1 | 2 | 2 | 2 | 6 |
| 300 | 0.5 | 0.65 | 1.89 | 0.54 | 1 | 2 | 2 | 2 | 4 |
| 1000 | 0.1 | 0.9 | 7.90 | 1.13 | 5 | 7 | 8 | 9 | 12 |
| 1000 | 0.1 | 0.8 | 5.85 | 0.75 | 4 | 5 | 6 | 6 | 9 |
| 1000 | 0.1 | 0.65 | 5.16 | 0.67 | 4 | 5 | 5 | 6 | 7 |
| 1000 | 0.35 | 0.9 | 4.62 | 1.16 | 3 | 4 | 5 | 5 | 9 |
| 1000 | 0.35 | 0.8 | 3.46 | 0.66 | 2 | 3 | 3 | 4 | 6 |
| 1000 | 0.35 | 0.65 | 2.86 | 0.56 | 2 | 3 | 3 | 3 | 5 |
| 1000 | 0.5 | 0.9 | 3.82 | 1.11 | 2 | 3 | 4 | 5 | 7 |
| 1000 | 0.5 | 0.8 | 2.67 | 0.66 | 2 | 2 | 3 | 3 | 6 |
| 1000 | 0.5 | 0.65 | 2.19 | 0.41 | 2 | 2 | 2 | 2 | 4 |
| 3000 | 0.1 | 0.9 | 10.35 | 1.28 | 8 | 9 | 10 | 11 | 15 |
| 3000 | 0.1 | 0.8 | 7.05 | 0.76 | 6 | 7 | 7 | 7 | 10 |
| 3000 | 0.1 | 0.65 | 6.01 | 0.70 | 5 | 6 | 6 | 6 | 9 |
| 3000 | 0.35 | 0.9 | 7.07 | 1.21 | 4 | 6 | 7 | 8 | 11 |
| 3000 | 0.35 | 0.8 | 4.33 | 0.74 | 3 | 4 | 4 | 5 | 7 |
| 3000 | 0.35 | 0.65 | 3.30 | 0.49 | 3 | 3 | 3 | 4 | 5 |
| 3000 | 0.5 | 0.9 | 6.31 | 1.98 | 3 | 5 | 6 | 7 | 40 |
| 3000 | 0.5 | 0.8 | 3.62 | 0.72 | 2 | 3 | 4 | 4 | 7 |
| 3000 | 0.5 | 0.65 | 2.59 | 0.62 | 2 | 2 | 3 | 3 | 5 |

In summary, when the cross-factor effects are moderate or large, and the decay of the random variation in the process for the latents is moderate or fast, very few iterations would be needed before two distinct factors effectively become indistinguishable. With modest cross-factor effects, and slower rates of decay, it may be continue to be possible to distinguish distinct factors especially if sample sizes are large and the process by which the underlying latent variables causally affect one other has not proceeded for a longer period of time.

## 6. Some Generalizations

We will now consider some generalizations of the results above, and show that similar phenomena may arise with an arbitrary set of *k* factors. A set of *k* causally related factors can likewise, in a one-wave factor analysis, give rise to patterns of association among item responses that, in equilibrium, as $t \to \infty$, suggest one factor is sufficient. However, in other cases, a set of $k$ causally related factors may give to patterns of association among item responses that, in equilibrium, suggest that more than one factor, but fewer than $k$ factors, are present.

Suppose once again that $Y^t = (Y_1^t, \ldots, Y_p^t)$ denotes a set of item responses measured at time $t$, and $\eta^t = (\eta_1^t, \ldots, \eta_m^t)$ is a set of latent factors at time $t$, with the standard factor analytic model with independent errors given by:

$$Y^t = \Lambda \eta^t + \varepsilon$$

and suppose that the latent factors $\eta^t$ may be causally related to each other so that:

$$\eta^t = B\eta^{t-1} + W^t.$$

Under the notation and assumptions of Theorem 1, we have that, as $t \to \infty$, $Y^t$ will converge in distribution to the random variable

$$Y^* = \Lambda^* \eta^* + \varepsilon$$

where, by Theorem 1, the dimensionality of $\eta^*$ will be rank($D^*$) and where $D^*$ is of the form:

$$D^* = \lim_{t\to\infty} D^t = \begin{pmatrix} I & 0 \\ 0 & 0 \end{pmatrix}$$

with $D$ being a matrix of Jordan normal form in the Jordan decomposition of $B$.

We can need to consider the rank of where $D^*$. Let $\sim$ denote an equivalence relation over $\{1, \ldots, m\}$ defined by $i \sim j$ if and only if there exists $\{k_1, \ldots, k_v\} \subseteq \{1, \ldots, m\}$ such that $k_1 = i, k_v = j$ and for all $r \in \{1, \ldots, v\}$ either $B_{k_r k_{r-1}} \neq 0$ or $B_{k_{r-1} k_r} \neq 0$. The distinct equivalence classes of the factors indexed by $\{1, \ldots, m\}$ are such that no factor in one equivalence class is causally related to any factor in any other equivalence class. By permuting indices, one can put the matrix $B$ in block matrix form

$$B = \begin{pmatrix} B_1 & \cdots & 0 \\ \vdots & \ddots & \vdots \\ 0 & \cdots & B_q \end{pmatrix}$$

where each $B_r$ corresponds to the causal relationships among the factors in the $r$th equivalence class. We then have that

$$B^t = \begin{pmatrix} B_1^t & \cdots & 0 \\ \vdots & \ddots & \vdots \\ 0 & \cdots & B_q^t \end{pmatrix}.$$

For each $B_r$ we can consider the Jordan decomposition $B_r = QD_rQ^{-1}$ for some invertible matrix $Q$ and with $D_r$ in Jordan normal form; we then have $B_r^t = QD_r^tQ^{-1}$ and

$$D^* = \lim_{t\to\infty} D^t = \begin{pmatrix} D_1^t & \cdots & 0 \\ \vdots & \ddots & \vdots \\ 0 & \cdots & D_q^t \end{pmatrix}$$

For each equivalence class, $r$, that constitutes only one factor, we have $B_r = D_r = (1)$, and this will contribute 1 to the rank of $D^*$. For each equivalence class, $r$, constituted by two factors, we have by the argument in the Corollary to Theorem 1, that this will likewise contribute at most 1 to the rank of $D^*$since $D_r^t$ can only converge and have rank 2, if $D_r$ itself is a 2×2 identity matrix, in which case $B_r$ would be a 2×2 identity matrix, but then each of the two factors would constitute its own equivalence class, contrary to the supposition that the two formed a single equivalence class. Thus, any equivalence class, $r$, that is constituted by only two factors, will at most contribute only 1 to the rank of $D^*$, and will thus, in equilibrium, as $t \to \infty$, reduce to at most a single factor. Finally, consider an equivalence class constituted by more than 2 factors. Suppose that $B_r$ is such that all its entries are strictly positive. It would then follow from the Perron–Frobenius theorem (Perron, 1907; Frobenius, 1912; Meyer, 2000) that there is a unique eigenvalue with largest absolute value that has an associated eigenspace of dimension 1. By Theorem 1 of Oldenburger (1940), for $D_r^t$ to converge as $t \to \infty$, all eigenvalues must be less than or equal to 1 in absolute value. With a unique eigenvalue with largest absolute value and with an eigenspace of dimension 1, it would follow that the Jordan normal form matrix $D_r$ would have at most a single entry of 1 on its diagonal, and by Theorem 1 of Oldenburger (1940), for $lim_{t\to\infty} D_r^t$ to converge, it would have dimension at most 1. Thus, for an equivalence class of latent factors wherein each factor in that equivalence class positively causally affects the others, this equivalence class will contribute at most 1 to the rank of $D^*$ and, in equilibrium, will thus reduce to at most a single factor. For an equivalence class of latent factors such that some of the factors causally affect others, but not all causal relations are present within the class, or for which some factors might negatively affect others, so that the entries of $B_r$ are not strictly positive, the dimensionality of $lim_{t\to\infty} D_r^t$ may exceed 1. There has been some work on relaxing the strict positivity requirement of the Perron-Frobenius theorem (Noutsos, 2006) and thus, in some of these cases, the equivalence class may likewise only contribute 1 to the rank of $D^*$ and thus reduce, in equilibrium, to at most a single factor, but this is not always guaranteed by the present results.

Thus, in many cases, each equivalence class, as defined above, will give rise, in equilibrium, to at most a single factor. This will always be the case in the models above when a factor constitutes its own equivalence class, when an equivalence class has only two factors, or when an equivalence class is such that each of the factors positively affects each of the others. It may or may not be the case in other settings. However, as discussed further below, the final case of an equivalence class such that each of the factors positively affects each of the others may be relevant in settings with a series of indicators that are all closely related to a single construct but represent distinct facets of the construct with potentially different causal relationships to other outcomes. This may in fact be a relatively common setting. And once again, if this were the case, it would follow that, over time, this setting would become indistinguishable, in one-way factor analysis, from that of having only a single factor.

In settings in which it is known in advance that certain subgroups of factors cannot affect other subgroups of factors, the analysis above would still be applicable within each subgroup of factors. In cases in which the factor structure might differ according to subgroups defined by some other variable, the analysis above would still be applicable within subgroups defined by that variable. However, for one of the most common uses of factor analysis – to attempt to establish the unidimensionality of a scale – the generalizations here are not in fact needed. A one-wave factor analysis suggesting evidence for a single factor could arise from a single underlying factor, or from two causally related factors, or from a set of $k$ causally related factors each of which positively affects the other. The central point, demonstrated already in Section 2, is that one cannot tell from evidence for unidimensionality from a one-wave factor analysis whether there is one factor or whether there may be more than one in the presence of causal effects of the factors upon one another.

## 7. Discussion

The implications of the present work for current psychometric practices are potentially far-reaching. Factor analysis with one wave of data seemingly cannot distinguish between factor models with a single factor versus those with two factors that causally affect one another over time. If causal relations between factors cannot be ruled out a priori, alleged empirical evidence from one-wave factor analysis for a single factor still leaves open the possibilities of a single factor or of two or more factors that causally affect one another. It would, moreover, as noted above, be very unusual for it to be the case that one were uncertain as to the dimensionality of a set of factors, but confident that, if there were several, they would be causally unrelated. In most cases, we thus effectively cannot distinguish between these possibilities when one-wave factor analyses conclude that only one factor is present. This arguably constitutes a substantial portion of factor analytic studies. In the models considered above, the conclusion of two factors from a one-wave factor analysis would suffice to conclude that one factor is insufficient; but the supposed conclusion of one factor does not preclude the possibility of two factors being present that are causally related.

The problems arise because of the inability to distinguish between causation and alleged conceptual relationships with one wave of data. It is well-known in statistics, and in the biomedical and social sciences, that correlation does not imply causation. The sub-discipline of causal inference (Pearl, 2009; Imbens and Rubin, 2015; Morgan and Winship, 2015; VanderWeele, 2015; Hernán and Robins, 2022) provides a formal framework to reason about the assumptions needed to move from conclusions of association to conclusions of causation. Such careful thought helps us avoid the fallacy that "Correlation implies causation." We might refer to this as the "causal fallacy." Unfortunately, however, a converse fallacy seems to typically arise in psychometric measurement evaluation, namely, that "Correlation cannot imply causation – it must indicate a conceptual relationship." This too, of course, is false. Correlation does not always indicate a causal relationship, but, sometimes, it does. Sometimes correlations arise from causal relationships. As shown in this paper, it is this second converse fallacy that underlies dimensionality assessment in most psychometric work on evaluating measures. We might refer to this converse fallacy that "Correlation cannot imply causation – it must indicate a conceptual relationship" as the "measurement fallacy." From the discussion above, it is arguably the case that this measurement fallacy in fact ought to be treated with the same level of critique and skepticism that is appropriately directed at the causal fallacy. Both are fallacies. Both fallacies need to be avoided. A fair amount of attention has been given to the causal fallacy. However, to date, the measurement fallacy has almost entirely been ignored. This needs to change.

The problems that the present paper makes clear may eventually require re-evaluation of a great deal of prior psychometric assessment of scales. It is important to note, however, that this does not necessarily imply that the scales themselves are problematic. Many of them may be reasonable assessments of a particular construct. However, what is problematic is not necessarily the scales themselves, but the evidence that has been used for claims of unidimensionality. As shown in this paper, distinct factors with causal relationships can, over time, seemingly collapse into a single factor. This may be especially problematic with items that, on the face of it, would seem to correspond to two or more distinct constructs with construct phenomena that may be causally related, and that are then claimed, from one-wave factor analysis, to be unidimensional. Such was the case with analyses above concerning anxiety and depression. However, such claims of unidimensionality may also be problematic with regard to a series of indicators that seem to correspond to a single construct, but that may constitute causally related, but distinct, facets. Indeed, the analysis concerning the generalizations given in Section 6 implies that even if each indicator represented its own

“factor”, if each of these positively affected the others, then over time the process will become indistinguishable from that corresponding to a single factor. The setting of indicators corresponding to distinct facets of a single construct with causal relationships between the different aspects of the phenomena represented by each indicator may in actuality be a very common setting, perhaps one in fact corresponding to most psychosocial phenomena. It is entirely possible that this insight has been regularly missed because of the causal relationships between distinct facets and the over-reliance on factor analysis with one wave of data. Furthermore, as discussed elsewhere (VanderWeele, 2022), even scales with an underlying univariate latent variable may still give rise to causal structures such that distinct indicators have differential causal effects on outcomes. Claims, therefore, based on one-wave factor analysis, that there is a single underlying univariate latent variable that is all that is causally relevant are thus highly problematic.

The way forward with regard to dimensionality assessment for a set of indicators is not entirely clear. It is clear that current practices are, in many contexts, flawed in the ways documented above. In the presence of potential causal effects amongst factors, almost certainly two waves of data collection on all item responses will be needed so as to attempt to disentangle causal from supposedly conceptual relationships. However, even with two waves of data available, further work would need to be done on the correct analytic approach. Exploratory structural equation modeling (Asparouhov and Muthén, 2009; Marsh *et al*., 2009) might provide a potential way forward as it allows for multiple waves of data, data-driven dimensionality assessment, and the specification of potential causal effects. It is possible that with two waves of data, if one were to impose time-invariant loadings and allow the factors at each wave to affect the other factors one wave later, but impose no other assumptions beyond standard linearity/normality, that this would suffice to correctly identify that, at wave 2, two factors independent of each other conditional on the past were present, with the two sets of items loading on separate factors. Intuitively, it might, in this way, be possible to use the associations between the wave 1 and wave 2 item responses to try to infer the causal relations among the factors, and then effectively use the correlations amongst item responses at wave 2, once the causal effects are ‘netted out’, to try to infer the underlying factor structure and factor loadings. This would, however, require further development and evaluation of the conditions under which such an approach would lead to correct identification of the underlying causal and factor analytic structure. Regardless of such future developments, however, the results of this paper suggest that we should be wary of claims of unidimensionality for psychosocial phenomena assessed by indicators and then evaluated by one-way factor analysis.

## Acknowledgements

This research was funded by the National Institutes of Health, U.S.A. The authors thank Bengt Muthén for helpful comments on the manuscript.